\documentclass{ifacconf2026}

\usepackage{graphicx}      
\usepackage{natbib}        
\usepackage{float} 

\begin{document}
\begin{frontmatter}

\title{Automating Cause–Effect Specification with Knowledge Graphs and Large Language Models} 

\thanks[footnoteinfo]{The authors gratefully acknowledge the financial support provided by the Department of Chemical Engineering at Imperial College London for this work}

\author{Javal Vyas},
\author{Milapji Singh Gill},
\author{Mehmet Mercangöz}

\address{Autonomous Industrial Systems Lab, Imperial College London, Imperial College Rd, South Kensington Campus, London, SW7 2AZ, United Kingdom}

\begin{abstract}
Engineering specifications such as interlocks, alarm 
rationalization tables, and cause-and-effect (C\&E) matrices 
remain central to process control and safety, yet their creation 
is still predominantly manual, document-driven, and prone to 
inconsistency. This paper presents a semantic--AI framework that 
automates the generation of C\&E logic by combining a knowledge 
graph (KG) with a constrained large language model (LLM) layer. 
The KG builds on an established modular alignment ontology to 
represent process structure, operating modes, faults, symptoms, 
causes, and mitigation actions in a machine-interpretable form. 
The LLM then transforms this information into operator-ready 
safety narratives and Semantic Web Rule Language (SWRL) rules 
under strict ontology and vocabulary constraints, grounding the generated artifacts in the underlying semantic model. The workflow is 
demonstrated on a modular process plant, showing how engineering 
semantics, diagnostic relations, and machine-verifiable 
specifications can be generated from a unified knowledge 
representation with reduced manual effort.
\end{abstract}


\end{frontmatter}
\section{Introduction}
Control logic, alarms, and safety instrumented functions form the
backbone of industrial automation. Engineering artifacts such as
interlock lists, alarm rationalization tables, and cause-and-effect
(C\&E) matrices serve as the contractual specification between design,
verification, and implementation (\cite{FernandezAdiego.2020}). Despite their critical role, these
artifacts are typically produced through manual interpretation of
P\&IDs, control narratives, and operational know-how
(\cite{Thambirajah.2009, FernandezAdiego.2020}). As a result, they often suffer from
inconsistent semantics, limited traceability, and high maintenance
overhead, particularly as process configurations change throughout the
engineering life cycle.

The increasing maturity of \textit{semantic technologies} and
\textit{large language models (LLMs)} offers a timely opportunity to
modernize this specification workflow. \textit{Knowledge graphs (KGs)},
grounded in engineering ontologies, provide a machine-interpretable
representation of process structure, operating modes, diagnostic
relations, and mitigation actions
(\cite{JohannesI.Single.2020, Gill.2024}). LLMs, in turn, have shown strong capabilities in transforming structured engineering information into both formal and natural-language artifacts, with applications ranging from FMEA support (\cite{Xia.24}) to flowsheet correction (\cite{Balhorn.2024}) and supervisory control in anomalous situations (\cite{Vyas.2025, GillVyas}). However, KGs alone cannot autonomously
derive specification logic, and free-form LLM outputs lack the
guarantees of formal syntax and verifiability required for
safety-critical automation.

This paper addresses this gap by proposing a five-layer
semantics-driven pipeline that turns heterogeneous engineering
data into a coherent C\&E specification. 
The pipeline produces three complementary, mutually consistent
artifacts from a single semantic source: a formal C\&E table, a set of
operator narratives, and a machine-verifiable SWRL rule base. The aim
is not to replace human expertise but to reduce manual effort and
ensure semantic coherence across these representations.

The remainder of this paper is structured as follows. Sec.~2 reviews
related work and outlines the research gap. Sec.~3 presents the
proposed pipeline in detail. Sec.~4 introduces the case study, and
Sec.~5 reports and discusses the results. Sec.~6 concludes the paper
and provides an outlook on future research.

\section{Related Work}

\subsection{Semantic Models for Process and Diagnostic Knowledge}
Ontologies and KGs are increasingly used in
process systems engineering to formalize equipment hierarchies,
operating modes, and diagnostic relations
(\cite{RUPPRECHT2026101209}). \cite{Hildebrandt.2020} introduced
modular, standards-based Ontology Design Patterns (ODPs) for CPS, providing reusable building
blocks for engineering knowledge. Building on this, alignment
ontologies grounded in established standards have been developed for
corrective maintenance of CPS
(\cite{Gill.2023,Westermann.2023}). At the application
level, ontology-based reasoning has been used to automate HAZOP
studies (\cite{JohannesI.Single.2020}). Earlier 
work has also leveraged structured information models (e.g., 
CAEX-based plant connectivity) for plant-wide cause--effect 
analysis (\cite{Thambirajah.2009}). More recently, the integration of KGs with LLMs has supported broader process-engineering tasks such as flowsheet
correction (\cite{Balhorn.2024}), P\&ID generation
(\cite{Gowaikar.17.12.2024}), and KG-driven task automation
(\cite{Sakhinana.2024}). Across these contributions, the focus has
been on static knowledge capture or on isolated assistance functions.
Limited effort has been directed toward generating executable or
machine-verifiable specification logic directly from the KG.

\subsection{LLM-Based Generation of Formal Rules and C\&E Logic}
Two research directions address the automation of formal specification
artifacts. The first uses LLMs to translate domain knowledge into
symbolic representations: \cite{Laurenzi.2024} show that LLMs can
generate SWRL rules in enterprise settings when guided by ontology
constraints, and \cite{Soularidis24} derive SWRL from domain texts,
although without process-engineering semantics or KG-grounded
context. Beyond rule generation, LLM agents have also been applied to
related verification-oriented tasks such as PLC code synthesis
(\cite{Agents4PLC}) and FMEA support (\cite{Xia.24}). 
The second targets C\&E matrices directly. C\&E tables remain
the primary mechanism for documenting safety and interlock logic, with methodologies such as the CERN CEM-based specification
(\cite{FernandezAdiego.2020}) providing rigorous semantics for manually authored matrices. Other automation efforts extract C\&E content
from P\&IDs, HAZOP worksheets, or alarm databases using rule- or
pattern-based techniques (\cite{Thambirajah.2009, JohannesI.Single.2020}), but rely on procedural heuristics that lack
formal semantics and are therefore difficult to verify or reuse. SWRL
in turn offers a reasoning-ready formalism that can be analyzed by
standard engines (e.g., Jess, Drools) for conflicts, redundancies, and
invariant violations, yet it is rarely used as a unifying
representation across system structure, diagnostic knowledge, and
mitigation logic.

\subsection{Research Gap}
Three observations emerge: 
\begin{enumerate}
    \item ontologies provide formal models of process structure and 
    diagnostics;
    \item LLMs can synthesize symbolic artifacts and natural-language 
    descriptions when constrained by such ontologies;
    \item C\&E generation itself remains largely manual, with limited 
    support for automated formal verification.
\end{enumerate}
To the authors' knowledge, no existing framework integrates these 
capabilities into a single workflow that automatically derives 
\emph{semantically grounded}, \emph{machine-verifiable} C\&E 
specifications from a process KG, while simultaneously generating 
operator-facing narratives and SWRL rules from the same semantic 
source. The present work addresses this gap by introducing a 
coherent pipeline.

\section{Pipeline}
\subsection{Overview}

\begin{figure*}[t]
    \centering
    \includegraphics[width=0.86\textwidth]{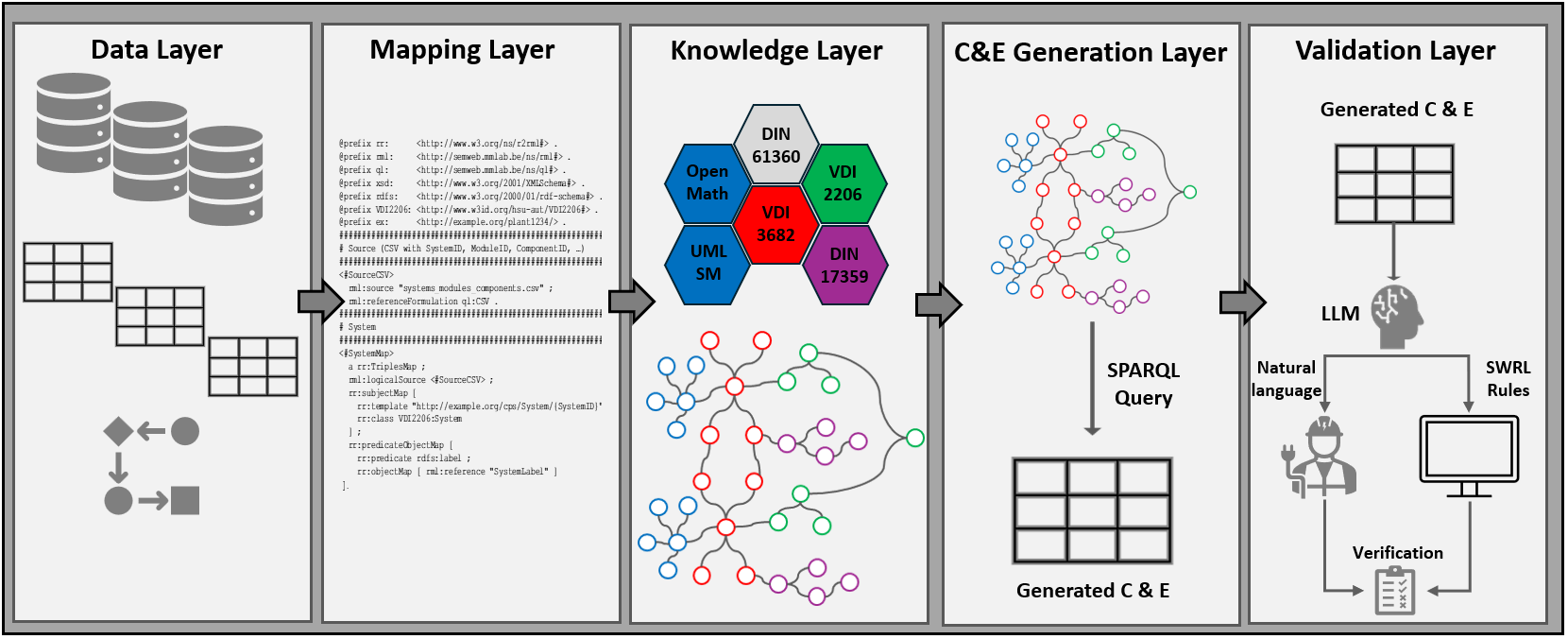}
    \caption{Overview of the proposed C\&E extraction pipeline.
    }

    \label{fig:pipeline}
\end{figure*}

The proposed pipeline for generating a C\&E representation 
integrates heterogeneous engineering data across five 
processing layers (see Fig.~\ref{fig:pipeline}). 

The \textit{Data 
Layer} aggregates distributed information from P\&IDs, process 
documentation, PLC and control logic, simulation models, and 
equipment specifications. These artifacts jointly capture the 
functional, structural, behavioral, and diagnostic knowledge 
required to relate observable effects to underlying causes.

In the \textit{Mapping Layer}, the artifacts are semantically lifted 
into RDF representations consistent with the alignment ontology, 
using RML, R2RML, or custom mapping pipelines. This step normalizes 
the heterogeneous sources into a unified semantic format.

The resulting RDF data is consolidated in the \textit{Knowledge 
Layer}, where the \texttt{CPSMod} alignment ontology integrates 
relevant knowledge based 
on established engineering and maintenance standards. The KG 
constructed in this layer serves as the semantic backbone of the 
pipeline.

The \textit{C\&E Generation Layer} derives a structured 
cause--effect representation from the KG through SPARQL queries 
that materialize C\&E rows from the diagnostic relationships 
encoded in the ontology. The resulting C\&E table is structured according to the ontology, which reduces ambiguities common in manually authored 
specifications.

Finally, the \textit{Validation Layer} translates the formal C\&E 
representation into operator-facing narratives and SWRL rules using 
a constrained LLM, and provides a human-in-the-loop review step in 
which domain experts inspect the generated artifacts and refine 
mappings or ontology assertions where necessary.

The remainder of this section focuses on the two layers that 
constitute the principal methodological contribution: the alignment 
ontology underlying the Knowledge Layer (Sec.~\ref{ont}), and the 
methodology for C\&E generation and validation 
(Sec.~\ref{sec:method_ce}). The Data and Mapping Layers follow 
established practices for ontology-based data integration and are 
not described in further detail.

\subsection{Alignment-Ontology} \label{ont}

C\&E modeling in complex process plants requires a knowledge representation that goes beyond simple fault\--symptom associations. To explain why certain effects occur in an engineered system, it is necessary to integrate multiple forms of knowledge: how the system is functionally organized, how its components are structurally linked, how it behaves over time, and which physical laws govern its operation. A cause-and-effect model must therefore capture causal dependencies across several interconnected dimensions. Functional dependencies describe how process steps influence downstream operations. Structural relationships reveal how components interact and constrain one another. Behavioral dynamics determine how system states and transitions mediate effects and physical relations explain how changes in variables such as flow, level, or pressure propagate through the system. Diagnostic knowledge, finally, provides the vocabulary for relating observable deviations to underlying fault mechanisms.

Such a multi-layered understanding of causation cannot be obtained from a single data artifact or modeling notation. Instead, it requires an integrated semantic representation that combines such knowledge into a coherent semantic model. To achieve this integration, we adopt an ontology-driven approach based on a set of ODPs aligned with established engineering and maintenance standards. These patterns provide reusable semantic building blocks that ensure conceptual consistency while allowing system knowledge to be formalized in a modular and extensible manner. In the context of this study, such an approach provides the foundation for deriving C\&E representations in a principled manner: causal pathways can be extracted consistently from the ontology, and the resulting models reflect the underlying engineering semantics rather than relying on informal or heuristic interpretations.

\begin{figure*}
    \centering
    \includegraphics[width=1.0\textwidth]{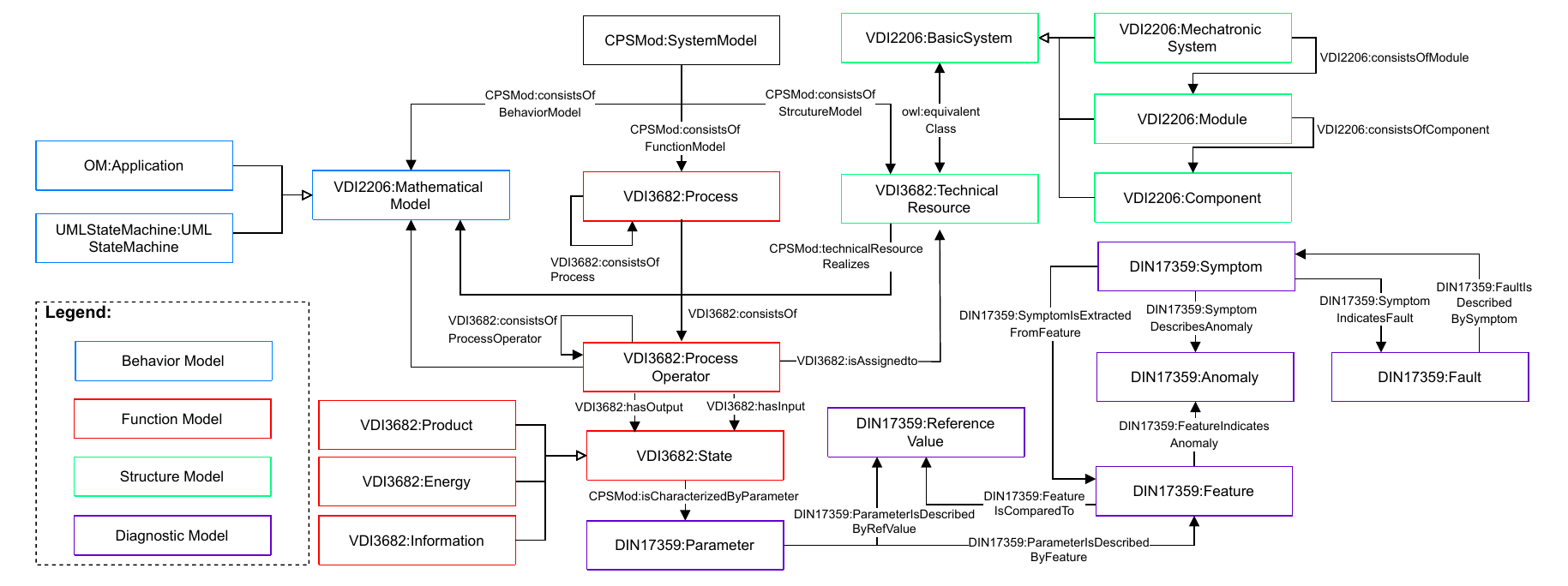}
    \caption{Reused alignment ontology showing how process, structural, behavioural,
condition, and diagnostic concepts are combined into a unified system model. }
    \label{fig:alignment}
\end{figure*}

The alignment ontology \texttt{CPSMod} (see Fig. \ref{fig:alignment}) used in this work incorporates several complementary ODPs with regard to the C\&E-Diagram. This alignment has been successfully applied in prior work on anomaly interpretation and knowledge-based diagnosis (\cite{Gill.2023, Gill.9920259122025}). Functional concepts, such as those derived from ODP \texttt{VDI 3682}, capture the whole process, separate process operators as well as their input and output products, energies and information, enabling the representation of causal relations along the operational workflow. Structural concepts, based on ODP \texttt{VDI 2206}, provide the system–module–component hierarchy, which locates causal connections within the physical architecture and clarifies how failures can propagate across components. Diagnostic concepts following ODP \texttt{DIN 17359} formalize the relations between features, reference values, symptoms, and faults, thereby supporting the mapping between observable effects and potential causes.

To model both discrete and continuous behavior of the system, the alignment ontology integrates two complementary behavioral ODPs. Discrete logic is represented using the ODP \texttt{UML State Machine}, which provides the state machine with all states, transitions, and events. These constructs express state-dependent behavior through guard conditions and state-specific actions, enabling effects that arise only in particular operational modes to be captured explicitly. Continuous dynamics are described using the \texttt{OpenMath} ODP, whose classes Application, Operator, and ObjectList allow mathematical relations to be encoded in structured form (\cite{Gill.9920259122025}). Specific operators are referenced via \texttt{OpenMath} content dictionaries (e.g.\ \texttt{calculus1} with \texttt{diff} for differentiation), so that complex behavior models can be represented as \texttt{OpenMath} applications linking actuator inputs to the evolution of measurable outputs.
Both behavioral layers are anchored in the functional model of ODP \texttt{VDI 3682}. A process operator (e.g., a filling step) references an associated state machine for its discrete modes or attaches relevant \texttt{OpenMath} equations for its continuous dynamics. Process operator inputs and outputs are thereby connected to both state transitions and physical equations, yielding coherent C\&E relations from manipulated variables to observable system responses.

\subsection{Methodology for C\&E Generation and Validation}
\label{sec:method_ce}

The methodology derives a machine-verifiable C\&E representation from the KG. In detail the C\&E generation happens with the help of the \textit{Knowledge Layer}, the \textit{C\&E Generation Layer}, and the \textit{Validation Layer}. Especially the following steps are relevant:
(i) semantic modeling of diagnostic knowledge using the fault class, 
(ii) automated materialization of \texttt{CauseEffectRow} individuals through SPARQL queries, and 
(iii) transformation of these rows into human-readable narratives and SWRL rules using an LLM.
Each step is detailed in the following.

\textbf{(i) Semantic Modelling of Faults, Causes, Symptoms, and Actions:} 
Rather than manually defining faults for each component,
we use the \emph{fault class} from ODP DIN17359 to capture recurring
diagnostic patterns, including valve faults (partially open, stuck
closed, leakage), pump faults (derating, loss of prime, cavitation),
pipe faults (clogging, partial obstruction), and operational
anomalies (filling-time increase, emptying-time deviation). Each
template specifies canonical links among faults
(\texttt{DIN17359:Fault}), symptoms (\texttt{DIN17359:Symptom}),
causes (\texttt{DIN17359:Cause}), and mitigation actions (instances
of a \texttt{:SafetyAction} hierarchy). Templates are instantiated
for concrete equipment items (e.g.\ inlet valves, pumps, bottling
lines) through SPARQL UPDATE statements within the
\textit{Knowledge Layer}.

\textbf{(ii) SPARQL-Based Construction of \texttt{CauseEffectRow} Instances:} The core of the method is the automatic construction of a formal C\&E representation
inside the KG. Each row is represented as an individual of class \texttt{:CauseEffectRow},
which connects the diagnostic and mitigation semantics as follows:

\begin{verbatim}
:CauseEffectRow
    :hasSubject  (equipment item)
    :hasFault    (fault)
    :hasSymptom  (symptom)
    :hasCause    (cause)
    :hasAction   (safety action)
    :hasMode     (operating mode) .
\end{verbatim}

These rows are not authored manually. Instead, a SPARQL CONSTRUCT rule traverses 
the diagnostic graph:

\begin{itemize}
    \item \texttt{:SafetyAction} individuals are taken as entry points;
    \item \texttt{:mitigatesFault} links actions to the associated faults;
    \item \texttt{DIN17359:DiagnosticSubjectHasFault} identifies the affected equipment item;
    \item \texttt{DIN17359:FaultIs\-Described\-BySymptom} attaches symptoms;
    \item \texttt{DIN17359:CausesFault} attaches root causes;
    \item \texttt{:occursInMode} attaches operating modes 
          (restricted to individuals of class \texttt{:OperatingMode}).
\end{itemize}

A unique identifier is generated for each row, and the result is 
inserted into the KG. By construction, the resulting C\&E 
representation:
\begin{itemize}
    \item is \emph{traceable} to the underlying engineering semantics,
    \item is \emph{consistent} with the diagnostic and structural dependencies encoded in the KG,
    \item contains \emph{no manually authored C\&E logic} beyond the templates.
\end{itemize}

\textbf{(iii) Tabular Export, Narrative, and SWRL Generation:}
For compatibility with engineering workflows, a SPARQL \texttt{SELECT} 
query extracts the C\&E rows into a tabular structure with columns 
for diagnostic subject, fault, symptom, cause, action, and mode. 
Labels are resolved using \texttt{rdfs:label} or local-name 
extraction. Each row is then passed to an LLM under prompt 
constraints that prohibit the invention of new thresholds, tags, or 
equipment. The LLM produces two complementary outputs per row: a 
human-readable safety narrative describing the fault condition and 
the associated mitigation, and a corresponding SWRL rule that 
encodes the same C\&E logic in a machine-processable form, 
referencing only classes, properties, and individuals defined in 
the ontology. The LLM's role is therefore \emph{interpretive} 
rather than generative: it reformulates the formally defined C\&E 
logic into operator-facing text and into rule syntax while 
preserving the semantics encoded in the KG.

\section{Case Study}
\begin{figure}[!b]
    \centering
    \includegraphics[width=\columnwidth]{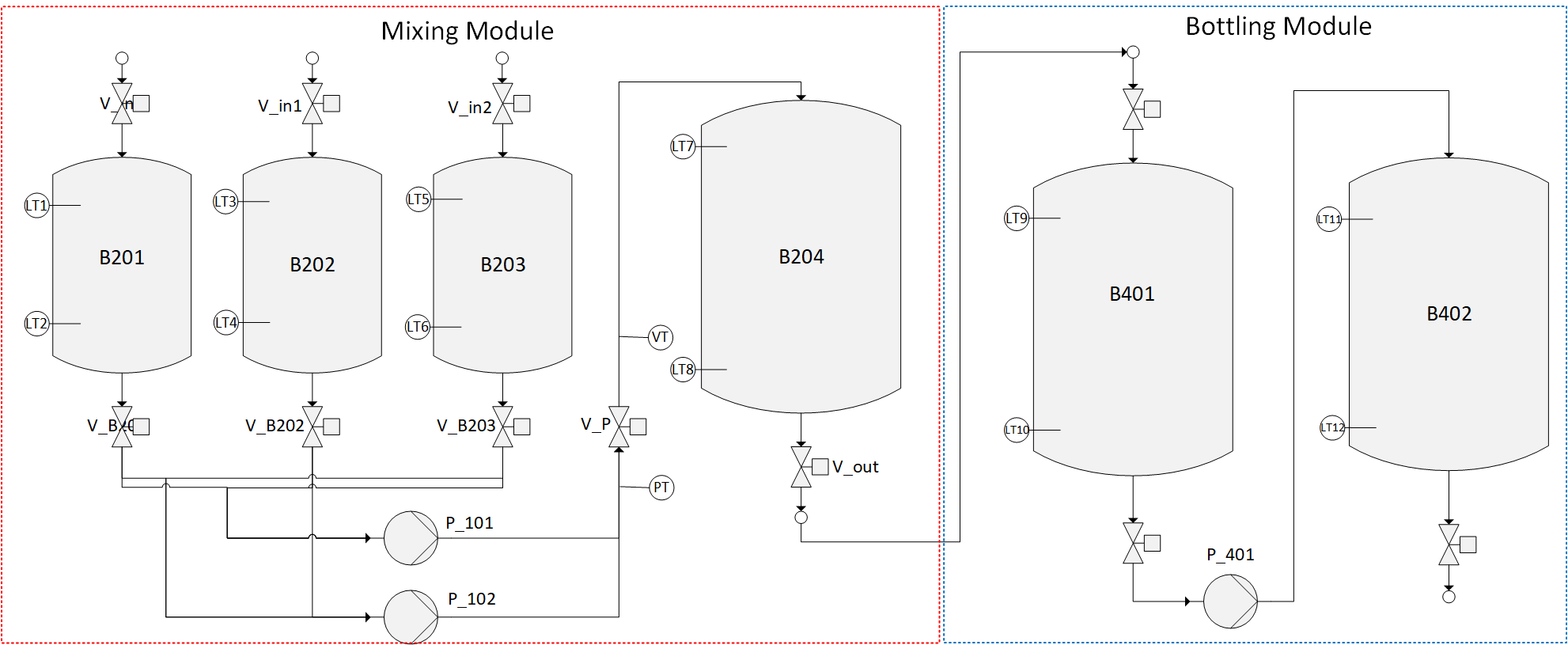}
    \caption{Modular process plant used as the case study in this work, 
showing the main interconnected processing modules and shared utilities.}
    \label{fig:plant}
\end{figure}

The case study is based on a modular process plant composed of two 
interconnected production modules: a mixing module (B201--B204) for 
sequential filling, agitation, and transfer operations, and a 
bottling module (B401--B402) for buffering and dosing 
(see Fig.~\ref{fig:plant}). Both modules consist of tanks, pumps, 
pipes, and valves that together implement the material flow from 
raw-material intake to final product dispensing.

To support the evaluation, several complementary data sources are 
available: a simulation model providing physical and control-logic 
behavior for nominal and faulty conditions, operational process data 
from real or emulated plant runs (sensor readings, actuator states, 
time-stamped events) stored in a relational database, and fault 
annotations in CSV format covering labels such as clogging, valve 
malfunction, and pump degradation. These datasets serve as input for 
the ontology-based integration and diagnostic reasoning steps.

\section{Results and Discussion}
\label{sec:results}

The evaluation considers three aspects of the workflow: extraction 
of C\&E relationships from the ontology, narrative generation, and 
SWRL synthesis. Each is discussed in turn before the overall 
findings are summarized.

\subsection{C\&E Extraction from the Knowledge Graph}

C\&E relationships were extracted directly from the ontology using a
SPARQL-based retrieval pipeline. The results are summarized in
Table~\ref{tab:metrics-overview}. A total of 15 C\&E rows were obtained,
covering all 8 distinct faults and all 8 diagnostic subjects contained
in the KG population. Six unique mitigation actions were identified,
including valve closures, pump trips, and fault-specific alarms. The
average actions-per-fault ratio of 1.00 indicates complete mitigation
coverage for all faults represented in the model.

\begin{table}[ht]
\centering
\caption{C\&E extraction and SWRL verification metrics.}
\label{tab:metrics-overview}
\begin{tabular}{l c}
\hline
\textbf{Metric} & \textbf{Value} \\
\hline
Total C\&E rows extracted from KG & 15 \\
Distinct faults represented & 8 \\
Distinct diagnostic subjects & 8 \\
Distinct mitigating actions & 6 \\
Average actions per fault & 1.00 \\
Semantic grounding rate (KG entities) & 100\% \\
SWRL rules syntactically valid & 100\% \\
Detected action conflicts & 0 \\
Unreachable or redundant rules & 0 \\
\hline
\end{tabular}
\end{table}

These results confirm that the alignment ontology (\texttt{CPSMod} + \texttt{DIN~17359} +
\texttt{VDI~3682}) provides a coherent and functionally complete representation of
faults, symptoms, operating modes, and mitigation logic. Each extracted
row corresponds to a meaningful causal pathway of the form
\emph{Fault~$\rightarrow$Subject~$\rightarrow$Mitigating Action}, which
forms the foundation for the subsequent LLM and SWRL synthesis steps.

\subsection{LLM-Generated Operator Narratives}

For each C\&E row, \texttt{gpt-4o-mini} generated a textual
explanation describing the fault condition and the associated
mitigation action. Because the model receives only the structured
entities from the KG (fault label, subject, and required action), the
generated narratives remain grounded in the ontology and do not invent
new causal relationships.

A qualitative evaluation of the narratives showed that they accurately
describe both the triggering condition and the intended operator
response. Alarms were expressed in operational terms
(e.g.\ ``inlet valve partially open leading to increased filling time''),
and shutdown actions were explained as protective safety interventions.
No hallucinated equipment or spurious failure modes were produced,
demonstrating the effectiveness of constraining the LLM with KG-derived
context. The narratives therefore serve as a readable companion to the
formal C\&E table, supporting operator understanding and documentation
workflows without compromising semantic correctness.

\subsection{SWRL Rule Synthesis and Verification}

The final step translated each C\&E row into a machine-verifiable SWRL
rule. Every generated rule adhered to SWRL syntax and referenced only
entities defined in the ontology, yielding a semantic grounding rate of
100\%. Automated verification checks were performed to detect logical
conflicts, such as contradictory commands on the same equipment item or
rules with unsatisfiable antecedents. No conflicts, unreachable rules,
or redundant rule patterns were identified.

The SWRL layer therefore acts as a formal contract between the KG and
the final C\&E table. Because each rule is both human-interpretable and
machine-checkable, inconsistencies can be detected early in the
workflow, before control logic is implemented in PLC or SIS hardware.
This provides a level of validation that is typically unavailable in
manual C\&E table development.

\subsection{Discussion}
Overall, the results demonstrate that the proposed hybrid approach
successfully unifies three traditionally disconnected artifacts:
structured engineering knowledge, human-readable narratives, and
machine-verifiable SWRL specifications. Compared to classical workflows, where the matrix, 
its underlying risk rationale, and the operator-facing documentation 
are maintained as separate, manually synchronized artifacts, the 
proposed pipeline derives all three from a single semantic source. 
Consistency is therefore guaranteed by construction rather than 
enforced by review, and the diagnostic context (faults, symptoms, 
causes, modes) becomes part of the specification itself rather than 
an external annotation. The strong alignment across all three layers, 
with complete fault coverage, zero conflicts, and narrative 
consistency, suggests that the approach is suitable for AI-assisted 
specification generation in industrial automation projects.

\section{Summary and Outlook}
This paper presented a hybrid framework that integrates a KG, LLMs, 
and SWRL-based reasoning to generate semantically consistent and 
machine-verifiable C\&E specifications for process automation. 
Engineering knowledge encoded in the \texttt{CPSMod} alignment 
ontology is transformed into three complementary artifacts: a 
formally structured C\&E table, operator-oriented natural-language 
narratives, and a conflict-free set of SWRL rules. The case study 
showed full fault coverage, accurate narrative grounding, and 
complete syntactic and semantic validity of the generated rules, 
illustrating the potential of combining semantic modeling and 
generative AI for more reliable and auditable specification 
workflows.

Several avenues for future work remain open. First, expanding the ontology 
population to include additional process units and more complex 
fault-propagation pathways would allow the framework to be validated at 
larger scales. Second, integrating rule execution engines or temporal 
reasoners could enable simulation of SWRL-driven logic under dynamic 
operating conditions. Third, incorporating human feedback loops, for 
example through reinforcement learning or structured preference models, may 
further improve narrative clarity and reduce LLM-generated ambiguities. 
Finally, linking the generated specifications to downstream PLC or DCS 
implementation pipelines would close the engineering loop, enabling 
end-to-end traceability from high-level requirements to executable control 
logic. These extensions would further strengthen the role of hybrid 
knowledge- and AI-based methods in the design and verification of 
next-generation autonomous industrial systems.

\bibliography{cas-refs}
\end{document}